\documentclass[conference]{IEEEtran}
\IEEEoverridecommandlockouts
\usepackage{cite}
\usepackage{amsmath,amssymb,amsfonts}
\usepackage{algorithmic}
\usepackage{graphicx}
\usepackage{textcomp}
\usepackage{xcolor}
\usepackage{multirow}
\usepackage{bbding}
\usepackage{tabularx}
\usepackage{booktabs}
\usepackage{soul}
\usepackage{authblk}

\usepackage{hyperref} 
\hypersetup{colorlinks=true,linkcolor=blue,urlcolor=blue}

\def\BibTeX{{\rm B\kern-.05em{\sc i\kern-.025em b}\kern-.08em
    T\kern-.1667em\lower.7ex\hbox{E}\kern-.125emX}}

\begin{document}
\title{Uncertainty-Aware Explainable Recommendation with Large Language Models
\\


\thanks{\\ \dag ~Co-first authors. \\ $^{*}$ Corresponding authors.}
}

\author[1]{Yicui Peng\dag}
\author[1]{Hao Chen\dag}
\author[2]{Chingsheng Lin}
\author[3]{\\Guo Huang$^*$}
\author[1]{Jinrong Hu}
\author[4]{Hui Guo}
\author[5]{Bin Kong}
\author[6]{Shu Hu}
\author[1]{Xi Wu}
\author[7]{Xin Wang$^*$}

\affil[1]{Chengdu University of Information Technology, Chengdu, China}
\affil[2]{Tunghai University, Taiwan, China}
\affil[3]{Leshan Normal University, Leshan, China}
\affil[4]{University at Buffalo, SUNY, NY, USA}
\affil[5]{Meta, CA, USA}
\affil[6]{Purdue University, Indianapolis, USA}
\affil[7]{University at Albany, SUNY, NY, USA}

\maketitle

\begin{abstract}
Providing explanations within the recommendation system would boost user satisfaction and foster trust, especially by elaborating on the reasons for selecting recommended items tailored to the user. The predominant approach in this domain revolves around generating text-based explanations, with a notable emphasis on applying large language models (LLMs). However, refining LLMs for explainable recommendations proves impractical due to time constraints and computing resource limitations. As an alternative, the current approach involves training the prompt rather than the LLM. In this study, we developed a model that utilizes the ID vectors of user and item inputs as prompts for GPT-2. We employed a joint training mechanism within a multi-task learning framework to optimize both the recommendation task and explanation task. This strategy enables a more effective exploration of users' interests, improving recommendation effectiveness and user satisfaction. Through the experiments, our method achieving 1.59 DIV, 0.57 USR and 0.41 FCR on the Yelp, TripAdvisor and Amazon dataset respectively, demonstrates superior performance over four SOTA methods in terms of explainability evaluation metric. In addition, we identified that the proposed model is able to ensure stable textual quality on the three public datasets.

\end{abstract}


\section{Introduction}

\begin{figure}[htbp]
\centerline{\includegraphics[scale=0.46]{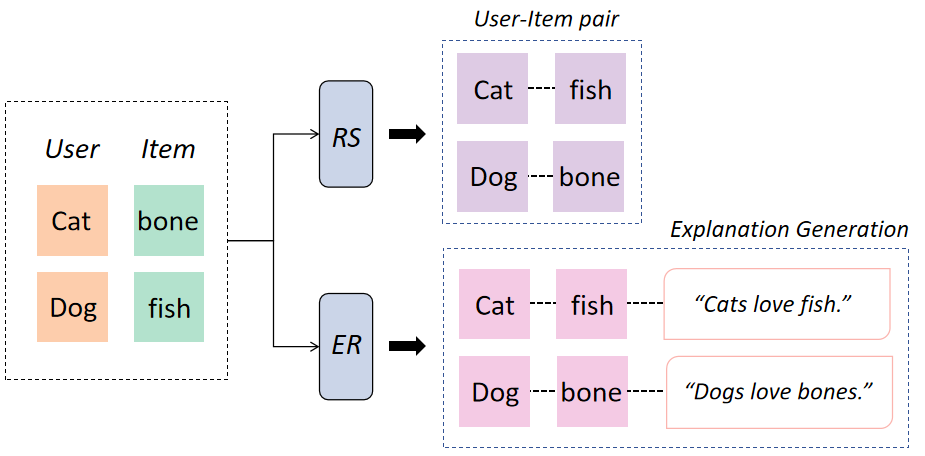}}
\caption{
RS and ER stand for Recommendation System and Explainable Recommendation respectively. 
Our goal is to leverage the power of LLM to generate natural language sentences that explain recommendations based on user-item pairs provided by RS.
}
\label{fig-view}
\end{figure}

The exponential growth in data volume overwhelms individuals due to challenges with managing and processing such vast amounts of information. The recommendation system (RS) \cite{palomares2021reciprocal} is a useful tool that analyzes historical data, such as user preferences and behavior to provide personalized suggestions or recommendations. These systems are commonly used in various domains, such as e-commerce, streaming services, and social media, to enhance user experience by offering relevant content, products, or services based on individual preferences \cite{lin2023predictive}. Early research in this domain focused on using logistic regression and collaborative filtering-based recommendations \cite{koren2009matrix, sarwar2001item, linden2003amazon}, \cite{zhang2014explicit}, to improve the accuracy of recommendation results. Later on, deep neural network models with advanced representational capacity facilitate modeling \cite{guo2017deepfm} greater diversity and complexity in the cross-interactions and combinations of features.

In recent years, the explainability of artificial intelligence (AI), particularly in RS has garnered significant attention from society including government, industry, and academic communities \cite{dong2019unified}. In this case, people not only depend on the results of recommendations but also need to be offered reasonable explanations for such recommendations in order to enhance their explainability.

Explainable recommendation \cite{zhang2020explainable,sun2020dual} refers to personalized recommendation algorithms that aim to address the "why" question. These algorithms offer both recommendation results and explanations, clarifying why specific items are recommended. For example, Fig.~\ref{fig-view}, this approach enhances transparency, persuasiveness, effectiveness, trustworthiness, and user satisfaction within RS.
Providing explanations for recommendations enhances the reliability of various data forms including videos, images and text \cite{zhang2014explicit}. The text-based generation of explanations has been widely studied, mainly attributed to the availability of textual data on online commercial platforms, like Yelp and TripAdvisor, which encourages users to express their opinions by writing reviews. Furthermore, this trend has also fostered advancements in natural language generation techniques, such as recurrent neural networks (RNN), transformer and pre-trained LMs \cite{dong2019unified, kenton2019bert, radford2018improving,liu2020sentence}.

In particular, LLMs have demonstrated their significant potential to be applied in understanding human expression, as well as in the explainability of recommendation. LLMs are pre-trained on massive amounts of data using self-supervised learning and have demonstrated remarkable success in learning universal representations \cite{gao2023chat}. 
Taking Generative Pre-Training (GPT) series \cite{radford2018improving, radford2019language, brown2020language} as an example, the initial GPT model \cite{radford2018improving} achieved state-of-the-art performance on 9 out of 12 natural language understanding tasks after fine-tuning. Subsequently, GPT-2 \cite{radford2019language} without fine-tuning demonstrated the ability to generate news articles resembling authentic ones. Further, GPT-3 \cite{brown2020language} could even do simple arithmetic that the model was not trained or fine-tuned for. However, the increasing model sizes and training data volumes present challenges, with GPT having 117 million parameters, while GPT-2 and GPT-3 significantly scaled up to 1.5 billion and 175 billion parameters, respectively.

Directly training a LLM from scratch for a recommendation task may require an enormous amount of task-specific data and computational resources. Current research has switched focus on training the prompts. The prompts for an LLM serve as a formulated input, akin to a question or instruction. This input guides the LLM during fine-tuning or specific task-oriented training, influencing the generated responses based on the model's learned patterns.

In the task of recommendation explanation generation, we would like to explore the reason why the item $i$ is suggested to the user $u$. 
However, the inclusion of user IDs and item IDs as prompts is challenging.
Simply providing IDs as isolated words to the language model (LM) does not enable the model to comprehend the implicit information conveyed. Li et al. \cite{li2023personalized} have illustrated the effectiveness of employing item features as distinct prompts to a pre-trained model for generating explanations. However, if we translate IDs into words as discrete prompts, there is a potential risk of losing the embedded feature information in the IDs. Taking the identification role of IDs as an example, the specific item can be represented by its characteristics words, such as $ i = \{ bathroom, subway, gym \} $ , while it cannot be asserted that these features must refer to the specific item. In this context, we are hardly possible to reverse the feature transformation back to the original IDs.
In addition, it is not necessary to limit the prompt to human-interpretable natural language (text-based content) \cite{liu2023pre}.
The purpose of constructing prompts is to find an effective approach that allows LLMs to accomplish the task rather than solely tailoring it to human comprehension.
The user and item can also be represented by vectors, either randomly initialized or produced by another model.
Specifically, input randomly initialized ID vectors as continuous prompts into LM for generating explanations, which is used in our approach.


In this paper, we introduce the recommendation as an auxiliary task in the explanation generation. This enables the transmission of user interests and item attributes to the generated explanations as comprehensively as possible.
Due to the complexity and time-consuming nature of manually adjusting the weights for two sub-tasks (rating prediction task and generative explanations task), we adopt a unified variance uncertainty approach to balance multiple loss functions \cite{kendall2018multi, liebel2018auxiliary, hu2023rank, hu2022sum, hu2020learning}.
But stochastic gradient descent is difficult to effectively optimize randomly initialized vectors, which can lead to sub-optimal results \cite{allen2019convergence}. To address this issue, we use a simultaneous optimization strategy.
Our contributions are summarized as follows:
\begin{itemize}
    \item  Harnessing the power of prompt learning for recommendation explanations: 
    Generate natural language explanations for recommendations using user and item IDs as continue prompts.
    The user and item are viewed as two special tokens for vectorized representation.
    \item  Dynamic learning of weights:
    The regularization term has been adjusted to enforce positive regularization values. 
    This enables the generated explanations to effectively convey user interests and item attributes, enhancing the overall quality of the recommendations.
\end{itemize}

In what follows, we first summarize related work in Section 2. We then propose prompt based learning for recommendation explainability and introduce the dynamic adjustment of loss functions. The details of our model are described in Section 3. The experimental setup is detailed in Section 4. The conclusion and future work are presented in Section 5.

\section{Related Work}
To address the problem of generating explanations, we adopt multi-task learning frameworks and prompt learning within the widely used LLMs.

\subsection{Explainable Recommendation}\label{2A}  
Improving the transparency of recommendations is a primary motivation for exploring explainable recommendations \cite{zhang2020explainable}. Beyond displaying the immediate result of a recommendation, it is crucial to provide users with meaningful explanations that elucidate why they are receiving a particular item. 
The interpretation of recommender systems takes different forms when faced with different practical application scenarios.
In general, explanations are typically categorized into two forms: feature-level explanations \cite{koren2009matrix, sarwar2001item, zhang2014explicit, pan2022accurate} and
text-based explanations, \cite{guo2017deepfm, liu2023pre, xian2021ex3, li2021personalized}. 
Feature-level explanations typically exploit the fact that a user may like an item feature to recommend it.
Past studies predominantly depend on creating text-based explanations as they are the most straightforward for users to comprehend and integrate into various application scenarios.
For example, Seo et al. \cite{seo2017interpretable} highlighted particular words and items in user reviews as explanations.
However, not all sentences in user reviews are for explanation purposes. Chen et al. \cite{chen2021generate} proposed a hierarchical sequence-to-sequence model that uses an auto-denoising mechanism based on topical item feature words for personalized explanation generation. Likewise, Li et al. \cite{li2021personalized} proposed a personalized transformer that generates aspect-inspired natural language explanations by using the IDs to predict the words in the target explanation. 
It not only predicts the user's preferences for the recommendations but also generates corresponding explanations.
Tan et al. \cite{tan2021counterfactual} designed a fixed template to formulate a joint optimization problem to generate recommendations.
While the aforementioned methods share the common goal of improving explainability, our approach distinguishes itself by considering user and item IDs. We assess the generated explanations not only based on textual quality but also from the perspective of explainability regarding item features.

\subsection{LLMs and Prompt Learning}\label{2B}
LMs are responsible for probabilistic modeling in the generation of natural language. 
When presented with a specific context, these models make predictions for the words that will be generated in context steps \cite{chen2023large}.
First, the majority of LMs are based on transformer or their related variants \cite{kenton2019bert, li2021personalized}.
Furthermore, the LMs undergo extensive pre-training using a massive amount of unlabeled corpus, followed by fine-tuning with task-specific data to ensure adaptability to diverse downstream applications \cite{liu2023pre}.
In recent years, there has been tremendous progress in LMs, and the emergence of LLMs like the GPT series \cite{radford2019language} marks a significant milestone for the entire AI community \cite{gao2023chat, chen2023large}.

LLMs are trained on vast amounts of text data to generate human-like sentences or paragraphs.
Which have proven to be highly effective in capturing complex linguistic patterns and dependencies. 
However, due to the wide variety of natural language tasks and the relatively small differences between them, fine-tuning an LM separately for each task is uneconomical.
One of the key aspects of LLMs is their ability to perform prompt learning \cite{liu2023pre}. 
In prompt learning, taking machine translations as an example, a template such as “Finnish: $[X]$ English: $[Z]$” is constructed first.
Then, the input placeholder is filled with a sample, for instance, "Finnish: Hyvää huomenta English: $[Z]$", which is referred to as a prompt.
With this approach, the model can be guided to predict the output placeholder $[Z]$ , such as "Good morning".

The rise of GPT series, especially GPT-3 \cite{brown2020language}, marked the beginning of prompt's popularization on natural language processing (NLP) tasks. 
Prompt learning refers to the process of fine-tuning pre-trained LMs with task-specific data, enabling them to adapt to various downstream applications \cite{lester2021power}.
This fine-tuning step plays a crucial role in enhancing the performance and applicability of LLMs in real-world scenarios.
Prompting templates can be classified into two categories based on the form of prompts: discrete prompts (a.k.a hard prompts) and continuous prompts (a.k.a soft prompts).
Discrete prompts are typically composed of natural language tokens that understandable by humans.
Brown et al. \cite{brown2020language} use manually designed prompts to direct GPT-3's generation. 
A typical prompt used in GPT-3 comprises a task description along with a few task-specific examples.
Li et al. \cite{li2023personalized} find some domain-specific words, such as movie titles and item features, which they adopted as prompts to represent the IDs. 
Then, these prompts were set to a fixed size and input into the LM to generate explanatory sentences.
However, Shin et al. \cite{shin2020autoprompt} point out that automatically generated textual prompts may lack explainability. 
Continuous prompts do not necessarily have to be words or even readable, which consist of a sequence of continuous vectors.
Li et al. \cite{li2021prefix} observe that such continuous prefix-based learning is more sensitive to different initialization in low-data settings than the use of discrete prompts with real words.
Lester et al. \cite{lester2021power} propose a simplified version of continuous prompts, which consists of virtual tokens that are only added to the embedding layer.
Thus, we prepend a sequence of continuous task-specific vectors as prompts to the input.

\subsection{Multi-task Learning in RS}\label{2C}
Multi-task learning aims to improve learning efficiency and prediction accuracy by learning multiple objectives from a shared representation \cite{kendall2018multi, lu2018like, wang2023deep}.
It has emerged as a powerful approach to jointly train models on multiple related tasks. 
Explainable recommender systems can be decomposed into recommendation tasks, which rating prediction, and explanation generation tasks that generate textual explanations for recommendations.
To overcome the limitations of single-task, researchers have shifted their attention towards multi-task learning in explainable recommender systems, where the generation of text explanation is specifically treated as an optimization task for learning.
It can achieve data augmentation through knowledge sharing among tasks. 
Additionally, it possesses advantages in feature recognition, computation, and storage. 
For example,
Li et al. \cite{li2017neural} used joint modeling of ratings and reviews beneficial for designing improved RS. 
They proposed a framework called NRT, which used deep neural networks to accurately predict ratings and simultaneously generate summary-style recommendation explanations with high linguistic quality.
However, it should be noted that this concise format of summary-style explanations may limit the amount of information conveyed.
Lu et al. \cite{lu2018like} proposed a model called MT for jointly learning rating prediction and recommendation explanation generation tasks. 
The MT uses Matrix Factorization (MF) \cite{koren2009matrix} for rating prediction and generative adversarial networks \cite{goodfellow2020generative} for generating explanatory texts.
Chen et al. \cite{chen2019co} delved into the explainability and accuracy of RS.
They introduced a joint-attention multi-task learning model called CAML, which takes into account the correlation between the recommendation task and the explanation task.

In this work, we focus on simultaneously optimizing the weights of the recommendation task and the generate explanations task.

\section{Methodology}
\begin{figure*}[tb]
\centerline{\includegraphics[scale=0.49]{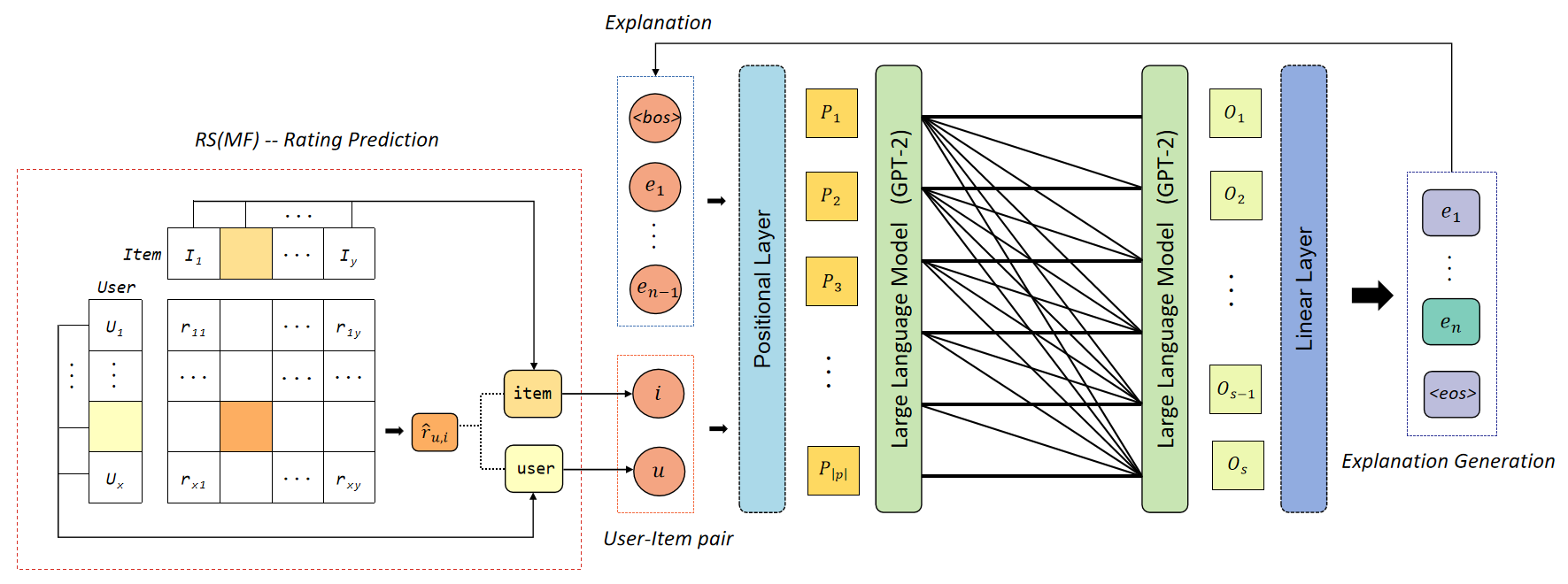}}
\caption{
The structure of multi-task learning involves the generation of a natural language sentence that explains why item $i$ is recommended to user $u$, which belongs to the Seq2Seq framework. 
Firstly, we use RS (MF) to obtain the rating prediction $\hat{r}_{u,i}$ , which is the result of the inner product between the user $u$ and item $i$. The loss function for rating prediction is calculated using the mean square deviation, as shown in \eqref{eq7}. Next, the user $u$ and item $i$ are treated as two special tokens for vectorization and serve as continuous prompts for one of the inputs, represented as $[u,i]$. Additionally, the other part of the input is the text of the recommendation explanation, denoted as $ [e_{1}, \cdots , e_{n-1}] $. The overall input is then passed through the LLM (GPT-2) and followed by a softmax fully connected layer. The output generates tokens sequence by sequence, starting from step 1 to each token's final representation $O_{s}$ for next-word prediction.
The objective function for optimization is the negative log-likelihood loss, as expressed in \eqref{eq4}. Finally, a joint training mechanism within a multi-task learning framework is employed to optimize the loss function, as described in \eqref{eq10}.
}
\label{fig-str}
\end{figure*}

In this section, we aim to formulate a coherent sentence explaining why an item $i$ is recommended to a user $u$, utilizing the user-item pairs $\mathcal{C}(u,i)$ provided by the RS. Throughout both training and testing phases, the input exclusively comprises user $u$ and item $i$, generating explanations as depicted in Fig.~\ref{fig-str}. In this section, we will describe prompt learning for explanation generation, introduce the rating prediction task and discuss the joint optimization strategy.

Before presenting our technical specifics, we provide a brief explanation of key notations that will be involved in our approach (see Tab.~\ref{tab-sym}). In addition, the token for both user and item represents a unique but meaningless term that corresponds to a specific user ID and item ID (for example, a user ID might be 'abc123user' and an item ID might be 'def456item'). The token for text sentence is a word (e.g.,``window'').
A discrete prompt is a word sequence, such as several item features, while a continuous prompt is a sequence of vectors, such as user and item embeddings in this work.

\begin{table}[ht]
\caption{Summary of the notations and concepts in this work} 
\begin{center}
\begin{tabular}{c|l}
\hline
\textbf{Symbol}& \textbf{Description}  \\
\hline
$\mathcal{U}$   &the sets of users in a dataset \\
$\mathcal{I}$   &the sets of items in a dataset \\
$\mathcal{C}$   &the set of user-item interactions in a RS  \\
$\mathcal{F}$   &the set of feature in a RS  \\
$\mathcal{E}$   &the set of all sequence of explanations in a RS  \\
$\mathcal{T}$   &training set \\
$\mathcal{V}$   &the set of word \\
$U$             &the embeddings of users \\
$I$             &the embeddings of items \\
$u$             &a user ID in a RS \\
$i$             &an item ID in a RS \\
$n$             &the length of the recommendation explanations \\
$z$             &the probability distribution over the vocabulary \\
$E$             &the sequence of explanations \\
$W$             &weight matrix \\
$b$             &weight vector \\
$\lambda$       &uncertainty weight \\
$\omega_{1} $   &rating prediction task \\
$\omega_{2} $   &generative explanations task \\
$\Theta$        &model parameters \\       
\hline
\end{tabular}   \label{tab-sym}
\end{center}
\end{table}

\subsection{Prompt Learning and Explanation Generation Task}\label{3A}

The pre-trained LLM undergoes extensive training on a diverse set of text data before being fine-tuned for specific tasks. Models like GPT-3 learn a wide range of language patterns and structures during this pre-training, enabling them to comprehend and generate human-like text. This pre-training process equips these models with the ability to understand contextual information and linguistic nuances, making them versatile for various NLP tasks without the need for extensive task-specific training. 

The foremost objective is to discover and devise suitable prompts that assist LLMs in better comprehending recommendation tasks and generating explanations.
We consider IDs as two distinct types of tokens, assigning the embeddings of users and items to the sets $U \in \mathbb{R}^{|\mathcal{U}| \times d}$ and $I \in \mathbb{R}^{|\mathcal{I}| \times d}$, respectively, where $|\mathcal{U}|$ and $|\mathcal{I}|$ represent the number of users and items in the dataset. The vector representation of one of the items is given by:
\begin{equation}
i=I^{T} g(i) 
\label{eq1}
\end{equation}
$g(i) \in \{0, 1\}^{|\mathcal{I}|}$ denotes a one-hot vector of the number of items in the dataset, with its non-zero element corresponding to the position of the item vector in $\mathcal{I}$. Likewise, the user $u$ representation can be obtained from $U$ using the same approach:
\begin{equation}
u=U^{T} g(u) 
\label{eq2}
\end{equation}

Conceptually, the input sequence is denoted as a prompt represented by $P = [u, i, E]$ to the LLMs (e.g.,GPT-2), where $u$ represents a user, $i$ represents an item. The representation of $(u,i)$ pair is followed by the sequence of the explanation which is denoted as $E = e_{1}, \cdots , e_{n}$. It consists of the words $e$ in explanation, where $|E|$ is the length of the sequence.

In the training stage, the initial randomization of user and item embeddings ($U$ and $I$) is succeeded by MF, enabling updates via back-propagation. This process leads to the generation of user representation ($u$) and item representation ($i$). Subsequently, the input representation ($P$) is formed by concatenating the user-item pair with the explanation, and an extra positional layer is introduced to capture the position of each token within the input sequence $P$. To be specific, the input is tokenized as $ [u, i, e_{1}, \cdots , e_{n-1}] $ and followed by the positional representations $[p_{1}, \cdots , p_{|P|}]$ where $|P|$ is the length of the sequence.
We denote the input representation of GPT-2 as $P  = [p_{1}, \cdots , p_{|P|}]$.
After passing $P$ through GPT-2 for producing the sequence's final $O  = [O_{1}, \cdots , O_{s}]$, where $|s|$ is the length of the output sequence. 
Then, we apply a linear layer with softmax function to each the sequence is obtained to map it onto a $\mathcal{V}$-sized vector.
After passing through this linear layer, the representation $O_{s}$ is turn into $z_{n}$ for next-word prediction.
\begin{equation}
z_{n} =softmax\left ( W_{O_{s} } +b \right ) 
\label{eq3}
\end{equation}
where $W \in \mathbb{R}^{|\mathcal{V}| \times d}$ , and $b \in \mathbb{R}^{|\mathcal{V}|}$ are weight parameters and $softmax(.)$ is the softmax function.
The value in vector $z_{n}$ stands for the probability of each token over the vocabulary $\mathcal{V}$.
For model learning, we adopt the objective function for optimization is the negative log-likelihood loss, and compute the mean of user-item pairs in the training set:
\begin{equation}
\mathcal{L}_{S}=\frac{1}{|\mathcal{T}|} \sum_{(u, i) \in \mathcal{T}} \frac{1}{\left|E\right|} \sum_{n=1}^{\left|E\right|}-\log z_{2+n}^{e_{n}}   
\label{eq4}
\end{equation}
where the value 2 in $-\log z_{2+n}^{e_{n}}$ represents the total number of the user ID and item ID, $|E|$ and $| \mathcal{T} |$ respectively denote the number of explanation words and the number of user-item pairs in the training set,
$z_{n}^{e_{n}} $ is the probability of generating the target token at step $n$.

At the stage of inference, our goal is to guide the model in generating an explanation word sequence ${E^{*}}$ that has the maximizes the log-likelihood.
\begin{equation}
E^{*}=\underset{E \in \hat{\mathcal{E}}}{\arg \max } \sum_{n}^{|E|} \log z_{2+n}^{e_{n}}
\label{eq5}
\end{equation}
There are several approaches to finding the sequence $E^{*}$ , such as greedy decoding and beam search. 
However, developing search algorithms is not our main focus, so we opt for a simple greedy decoding strategy. 
This entails selecting the word with the highest probability as the prediction at each step.
Initially, we provide the prompt $u$ and $i$, and input a special begin-of-sequence token $<bos>$ to the model. 
Based on the resulting probability distribution $z_{<bos>}$ for the next word, we select the word with the highest probability as our prediction. 
Subsequently, we append this predicted word to the end of the sequence, creating a new input sequence for generating the next word.
We repeat this process until the model generates a special token $<eos>$ or the generated explanation reaches a predefined length.

\subsection{Rating prediction Task}\label{3B}
Due to the fact that user ratings on items $r_{u,i}$ inherently imply a user-item relationship, e.g. representing the degree of user appreciation towards the items, they can serve as informative signals for learning to generate explanations.
We adopt the MF recommendation model to complement the generation of explanations.
Specifically, the predicted rating is derived from the dot product of the target user and the item:
\begin{equation}
\hat{r}_{u,i} = U^{T} i 
\label{eq6}
\end{equation}
Where $r_{u,i}$ is the ground-truth rating that user $u$ to item $i$. 
So the recommendation rating task loss function is calculated using the mean square error:
\begin{equation}
\mathcal{L}_{R} = \frac{1}{\left | \mathcal{T} \right | } \sum_{\left ( u,i \right ) \epsilon \mathcal{T}}  \left ( r_{u,i}-\hat{r}_{u,i} \right )^{2}  
\label{eq7}
\end{equation}

\subsection{Joint Optimization of Two Sub-tasks}\label{3C}
To fully leverage the RS for explanation generations, we integrate the rating prediction task and the generative explanation task into a multi-task learning framework.
However, the performance of the model is sensitive to the selection of weights, $\lambda _{S}$ and $\lambda _{R}$.
The process of tuning the weights of two tasks is complex and time-consuming by hand.
Multi-task learning is usually different tasks on the same dataset, so the weight of the loss function is measured by considering the homoskedastic uncertainty.
The overall loss function is formulated as:
\begin{equation}
\mathcal{L}\left ( \omega_{1}, \omega_{2} ,\lambda _{S},\lambda _{R}  \right ) =\lambda _{R}\mathcal{L}_{R}(\omega_{1})+\lambda _{S}\mathcal{L}_{S}(\omega_{2})
\label{eq8}
\end{equation}
where $\lambda _{S}$ and $\lambda _{R}$ are trainable weights with an initial value set to 1.

Kendall et al. \cite{kendall2018multi} have demonstrated the effectiveness of their multi-task learning framework and provided a detailed derivation and proof of an equation.
However, for the purpose of brevity, we present here only the final equation achieved using their approach:
\begin{equation}
\mathcal{L}\left ( \omega ,\lambda_{t} \right ) =
\sum_{{t} \in \{S,R\}}  \frac{1}{2 \lambda_{t}^{2}} \mathcal{L}_{t}(\mathrm{ \omega})+\log (\lambda_{t})
\label{eq9}
\end{equation}
where $\mathcal{L}_{t}(\omega)$ represents the loss function for task $t$ and $\lambda _{t}$ denotes the trainable weight associated with task $t$.
However, during the iterative process, we observed that $\lambda$ could potentially be a significantly negative value, resulting in a negative loss calculation in such cases.
To address this issue and prevent the training loss from becoming negative, a slight adjustment is made to the proposed equation. 
The final loss function is as follows:
\begin{equation}
\mathcal{L}\left ( \omega ,\lambda _{t}  \right ) =
\sum_{{t} \in \{S,R\}} \frac{1}{2 \lambda _{t}^{2}} \mathcal{L}_{t}(\mathrm{ \omega})+\log (1+\lambda _{t}^{2})
\label{eq10}
\end{equation}
The trainable weights $\lambda$ and model parameters are trained together, converging to obtain the optimal values for weight initialization.
This modification ensures that the training loss remains positive, addressing the issue of negative loss values.

\section{Experimental}

\subsection{Datasets and Evaluation Metrics}\label{4A}

We conduct experiments on three publicly available datasets  \cite{li2021personalized} are from Yelp (restaurant) \cite{yelp}, TripAdvisor (hotel) \cite{tripadvisor} and Amazon (movies and TV) \cite{amazon}. The sample in datasets consists of a user ID, an item ID, a rating on a scale of 1 to 5, an explanation deriving from users' reviews (e.g.,``the gym area had excellent facilities'') and an item feature (e.g.,``subway station''). 
Each explanation contains at least one item feature (e.g.,``gym''), which ensures the explanation quality and the ``\%Explanation" in the table indicates the percentage of  value of the feature word as a percentage of the sentence. The detailed statistics of three datasets are presented in Tab.~\ref{tab-dataset}.


In order to provide a comprehensive evaluation of our model and compare its results with the others' baseline, we adopt the same evaluation metrics as in previous work \cite{li2023personalized}, which can be divided into two aspects, explainability and textual quality. The former aims to investigate whether the model has capability to generate a diverse explanation containing useful information. The latter is a set of traditional textual matching metrics that are used for examining whether the generated sentences have a similar meaning with the explanation. To be specific, the explainability used Unique Sentence Ratio(USR) \cite{li2020generate}, Feature Coverage Ratio (FCR) \cite{li2020generate} and Feature Diversity (DIV) \cite{li2020generate}.

\begin{itemize}
    \item USR is used to quantify the proportion of distinct sentences generated among all samples. The definition is as follows \eqref{eq11}, where $| \varepsilon^{*} |$ represents the number of generated unique explanations, and $N$ is the total number of total explanations.
    And only the exactly matched sentences are considered identical.
    \begin{equation}
    USR=\frac{ \left | \varepsilon ^{*} \right | }{N} 
    \label{eq11}
    \end{equation}
    
    \item FCR calculates the number of distinct features exhibited in the generated explanations. The equation is shown in (\ref{eq12}), where $\mathcal{F}$ represents the collection of all features in the dataset, and $N_{g}$ denotes the count of distinct features shown in the generated explanations.
    \begin{equation}
    FCR=\frac{N_{g} }{\left | \mathcal{F} \right | } 
    \label{eq12}
    \end{equation}
    
    \item Following the FCR, the diversity of generated explanations is measured by DIV, which quantifies the extent of feature intersection between any two generated explanations $\hat{\mathcal{F}}_{u, i}$ and $\hat{\mathcal{F}}_{u^{\prime}, i^{\prime}}$ .
    \begin{equation}
    DIV=\frac{2}{N \times(N-1)} \sum_{u, i, u^{\prime},i^{\prime}} \left|\hat{\mathcal{F}}_{u, i} \cap \hat{\mathcal{F}}_{u^{\prime}, i^{\prime}}\right|
    \label{eq13}
    \end{equation}
\end{itemize}

The textual quality is evaluated using two traditional textual matching metrics, BLEU \cite{papineni2001bleu} (commonly used in machine translation) and ROUGE \cite{lin2004rouge} (commonly used in text summarization).
Where the key insight is to count the n-gram level overlaps between the model predicted explanation and real user reviews.

In general, DIV is assessed as ``the lower, the better," whereas all other metrics follow the principle of ``the higher, the better."

\begin{table}[t]
\caption{Statistics of the three datasets}
\centering
\scalebox{0.9}{
\begin{tabular}{@{}lcccccc@{}}
\toprule
Dataset      & \#Users  & \#Items  & \#Ratings  & \#Features   & \%Explanation  \\ \midrule
Yelp \cite{yelp}  & 27,147   & 20,266   & 1,293,247  & 7,340   & 12.32     \\
TripAdvisor \cite{tripadvisor}  & 9,765  & 6,280   & 320,023   & 5,069  & 13.01   \\
Amazon \cite{amazon}  & 7,506  & 7,360  & 441,783  & 5,399  & 14.14   \\ \bottomrule
\end{tabular}  }
\label{tab-dataset} 
\end{table}

\begin{table*}[t]
\caption{Performance comparison of all neural generation methods in terms of Personalization across three datasets}
\centering
\scalebox{1.05}{
\begin{tabular}{clllllllllll}
\hline
\multirow{2}{*}{Methods}   & \multicolumn{3}{|c|}{Explainability}     & \multicolumn{8}{c}{Textual Quality}     \\  \cline{2-12}
& \multicolumn{1}{|c}{DIV↓}   & \multicolumn{1}{c}{USR↑}  & \multicolumn{1}{c|}{FCR↑} 
& \multicolumn{1}{c}{BLEU-1↑}     & \multicolumn{1}{c|}{BLEU-4↑}   & \multicolumn{1}{c}{R1-Pre↑} 
& \multicolumn{1}{c}{R1-Rec↑}   & \multicolumn{1}{c|}{R1-F1↑} & \multicolumn{1}{c}{R2-Pre↑} 
& \multicolumn{1}{c}{R2-Rec↑}   & \multicolumn{1}{c}{R2-F1↑} \\  \hline
\hline
\multicolumn{12}{c}{Yelp}       \\ \hline
\multicolumn{1}{l|}{Attn2Seq \cite{dong2017learning} }   & \multicolumn{1}{c}{2.25}    & \multicolumn{1}{c}{0.05}  
& \multicolumn{1}{c|}{0.05}     & \multicolumn{1}{c}{10.25}   & \multicolumn{1}{c}{0.54}  
& \multicolumn{1}{c}{17.13}     & \multicolumn{1}{c}{11.44}   & \multicolumn{1}{c}{12.72} 
& \multicolumn{1}{c}{1.49}      & \multicolumn{1}{c}{1.13}    & \multicolumn{1}{c}{1.16}   \\ 
\multicolumn{1}{l|}{NRT \cite{li2017neural}  }      & \multicolumn{1}{c}{1.67}  & \multicolumn{1}{c}{0.20} 
& \multicolumn{1}{c|}{0.12}   & \multicolumn{1}{c}{10.92} & \multicolumn{1}{c}{0.60} 
& \multicolumn{1}{c}{16.73}   & \multicolumn{1}{c}{11.91} & \multicolumn{1}{c}{12.89} 
& \multicolumn{1}{c}{1.63}    & \multicolumn{1}{c}{1.21}  & \multicolumn{1}{c}{1.26}  \\ 
\multicolumn{1}{l|}{PETER \cite{li2021personalized}  }     & \multicolumn{1}{c}{1.62}   & \multicolumn{1}{c}{0.15} 
& \multicolumn{1}{c|}{0.15}    & \multicolumn{1}{c}{10.74}  & \multicolumn{1}{c}{0.63} 
& \multicolumn{1}{c}{16.18}    & \multicolumn{1}{c}{11.90}  & \multicolumn{1}{c}{12.63} 
& \multicolumn{1}{c}{1.60}     & \multicolumn{1}{c}{1.32}   & \multicolumn{1}{c}{1.28}  \\ 
\multicolumn{1}{l|}{PEPLER \cite{li2023personalized}}      & \multicolumn{1}{c}{1.66}    & \multicolumn{1}{c}{\textbf{0.30}} 
& \multicolumn{1}{c|}{0.27}      & \multicolumn{1}{c}{\textbf{11.70}}  & \multicolumn{1}{c}{\textbf{0.75}} 
& \multicolumn{1}{c}{17.52}      & \multicolumn{1}{c}{\textbf{12.85}}  & \multicolumn{1}{c}{\textbf{13.72}} 
& \multicolumn{1}{c}{1.86}       & \multicolumn{1}{c}{\textbf{1.45}}   & \multicolumn{1}{c}{\textbf{1.48}}  \\
\multicolumn{1}{l|}{OURS}     & \multicolumn{1}{c}{\textbf{1.59}}   & \multicolumn{1}{c}{\textbf{0.30}} 
& \multicolumn{1}{c|}{\textbf{0.34}}   & \multicolumn{1}{c}{11.10}  & \multicolumn{1}{c}{0.72} 
& \multicolumn{1}{c}{\textbf{17.88}}   & \multicolumn{1}{c}{12.49}  & \multicolumn{1}{c}{13.52} 
& \multicolumn{1}{c}{\textbf{1.89}}    & \multicolumn{1}{c}{1.38}   & \multicolumn{1}{c}{1.44}  \\ \hline
\hline
\multicolumn{12}{c}{TripAdvisor}        \\ \hline
\multicolumn{1}{l|}{Attn2Seq \cite{dong2017learning}}     & \multicolumn{1}{c}{4.74}      & \multicolumn{1}{c}{0.02} 
& \multicolumn{1}{c|}{0.05}       & \multicolumn{1}{c}{15.20}      & \multicolumn{1}{c}{0.96} 
& \multicolumn{1}{c}{18.74}      & \multicolumn{1}{c}{\textbf{16.42}}     & \multicolumn{1}{c}{16.38} 
& \multicolumn{1}{c}{2.42}       & \multicolumn{1}{c}{\textbf{2.32}}      & \multicolumn{1}{c}{2.19}   \\ 
\multicolumn{1}{l|}{NRT \cite{li2017neural}}       & \multicolumn{1}{c}{6.07}     & \multicolumn{1}{c}{0.01} 
& \multicolumn{1}{c|}{0.02}    & \multicolumn{1}{c}{13.76}    & \multicolumn{1}{c}{0.80} 
& \multicolumn{1}{c}{19.01}    & \multicolumn{1}{c}{14.57}    & \multicolumn{1}{c}{15.58} 
& \multicolumn{1}{c}{2.10}     & \multicolumn{1}{c}{1.59}     & \multicolumn{1}{c}{1.68}   \\ 
\multicolumn{1}{l|}{PETER \cite{li2021personalized} }       & \multicolumn{1}{c}{3.62}     & \multicolumn{1}{c}{0.05} 
& \multicolumn{1}{c|}{0.09}      & \multicolumn{1}{c}{15.13}    & \multicolumn{1}{c}{1.00} 
& \multicolumn{1}{c}{18.30}      & \multicolumn{1}{c}{16.15}    & \multicolumn{1}{c}{16.00} 
& \multicolumn{1}{c}{2.24}       & \multicolumn{1}{c}{2.23}     & \multicolumn{1}{c}{2.06}   \\ 
\multicolumn{1}{l|}{PEPLER \cite{li2023personalized}}       & \multicolumn{1}{c}{2.89}      & \multicolumn{1}{c}{0.21} 
& \multicolumn{1}{c|}{0.21}       & \multicolumn{1}{c}{\textbf{16.02}}     & \multicolumn{1}{c}{\textbf{1.15}} 
& \multicolumn{1}{c}{\textbf{19.52}} & \multicolumn{1}{c}{16.31}     & \multicolumn{1}{c}{\textbf{16.69}} 
& \multicolumn{1}{c}{\textbf{2.53}}  & \multicolumn{1}{c}{\textbf{2.32}}  & \multicolumn{1}{c}{\textbf{2.22}}   \\
\multicolumn{1}{l|}{OURS}      & \multicolumn{1}{c}{\textbf{1.92}}    & \multicolumn{1}{c}{\textbf{0.57}} 
& \multicolumn{1}{c|}{\textbf{0.50}}    & \multicolumn{1}{c}{15.33}    & \multicolumn{1}{c}{1.08} 
& \multicolumn{1}{c}{18.29}    & \multicolumn{1}{c}{15.62}    & \multicolumn{1}{c}{15.66} 
& \multicolumn{1}{c}{2.25}     & \multicolumn{1}{c}{2.07}     & \multicolumn{1}{c}{1.98}   \\ \hline
\hline
\multicolumn{12}{c}{Amazon}        \\ \hline
\multicolumn{1}{l|}{Attn2Seq \cite{dong2017learning}}   & \multicolumn{1}{c}{2.64}     & \multicolumn{1}{c}{0.05} 
& \multicolumn{1}{c|}{0.04}     & \multicolumn{1}{c}{12.07}    & \multicolumn{1}{c}{0.73} 
& \multicolumn{1}{c}{18.35}     & \multicolumn{1}{c}{12.86}    & \multicolumn{1}{c}{14.14} 
& \multicolumn{1}{c}{2.01}      & \multicolumn{1}{c}{1.56}     & \multicolumn{1}{c}{1.61} \\ 
\multicolumn{1}{l|}{NRT \cite{li2017neural}}      & \multicolumn{1}{c}{2.71}     & \multicolumn{1}{c}{0.09} 
& \multicolumn{1}{c|}{0.04}   & \multicolumn{1}{c}{12.06}    & \multicolumn{1}{c}{0.69} 
& \multicolumn{1}{c}{17.17}   & \multicolumn{1}{c}{13.15}    & \multicolumn{1}{c}{13.83}
& \multicolumn{1}{c}{1.94}    & \multicolumn{1}{c}{1.68}     & \multicolumn{1}{c}{1.64}   \\
\multicolumn{1}{l|}{PETER \cite{li2021personalized} }      & \multicolumn{1}{c}{2.16}     & \multicolumn{1}{c}{0.20}  
& \multicolumn{1}{c|}{0.09}     & \multicolumn{1}{c}{11.75}    & \multicolumn{1}{c}{0.89} 
& \multicolumn{1}{c}{16.51}     & \multicolumn{1}{c}{13.10}     & \multicolumn{1}{c}{13.55} 
& \multicolumn{1}{c}{1.96}      & \multicolumn{1}{c}{1.76}     & \multicolumn{1}{c}{1.68}   \\ 
\multicolumn{1}{l|}{PEPLER \cite{li2023personalized}}      & \multicolumn{1}{c}{2.18}     & \multicolumn{1}{c}{0.35} 
& \multicolumn{1}{c|}{0.24}      & \multicolumn{1}{c}{13.46}    & \multicolumn{1}{c}{1.02} 
& \multicolumn{1}{c}{18.30}      & \multicolumn{1}{c}{14.37}    & \multicolumn{1}{c}{14.92} 
& \multicolumn{1}{c}{2.29}       & \multicolumn{1}{c}{1.92}     & \multicolumn{1}{c}{1.90}    \\ 
\multicolumn{1}{l|}{OURS}    & \multicolumn{1}{c}{\textbf{1.95}}    & \multicolumn{1}{c}{\textbf{0.53}} 
& \multicolumn{1}{c|}{\textbf{0.41}}  & \multicolumn{1}{c}{\textbf{14.07}}    & \multicolumn{1}{c}{\textbf{1.28}} 
& \multicolumn{1}{c}{\textbf{18.43}}  & \multicolumn{1}{c}{\textbf{14.96}}    & \multicolumn{1}{c}{\textbf{15.28}} 
& \multicolumn{1}{c}{\textbf{2.75}}   & \multicolumn{1}{c}{\textbf{2.32}}     & \multicolumn{1}{c}{\textbf{2.30}}  \\ \hline
\multicolumn{10}{l}{$^{\mathrm{*}}$R1 and R2 stand for ROUGE-1 and ROUGE-2. 
Pre and Rec respectively represent precision and recall.}
\end{tabular} }
\label{tab3} 
\end{table*}

\subsection{Baseline and Experimental Settings}\label{4B}
To carry out the comparison experiments, we consider four state-of-the-art baseline methods as follows:

\begin{itemize}
\item  \textbf{Attn2Seq} \cite{dong2017learning}  
is a review generation approach where the explanations are treated as reviews.
In this model, an attribute module is used to represent input attributes as vectors.
Furthermore, it introduces a sequence decoder enhanced with attention, conditioned on these vectors, to generate reviews.

\item  \textbf{NRT} \cite{li2017neural}  
uses a GRU \cite{li2020generate}  model with context information to translate user and item latent factors into abstractive tips for generating explanations. 

\item  \textbf{PETER} \cite{li2021personalized}  
introduces present a PErsonalized Transformer for explainable recommendations.
They designed an additional task to establish a connection between IDs and words.
In addition to generating explanations, the model can also do rating predictions.

\item  \textbf{PEPLER} \cite{li2023personalized}  
is a simple and effective framework that attempts to use user and item IDs to generate explanations. 
This approach explores the potential of pre-trained Transformer models.

\end{itemize}

All of the above are implemented using the paper provided code.
Following previous works \cite{li2021personalized} , we randomly split each dataset into training, validation, and testing sets using an 8:1:1 ratio. 
Moreover, the splitting process was repeated five times and the final experimental results were reported as the average of these five experiments.
For model training, we used the training set and performed hyperparameter tuning on the validation set.
Subsequently, we evaluated the model's performance on the test set.
Notably, the training set is designed to include at least one record for every user and item in the dataset.

The proposed model uses the following default hyperparameters: Adam optimizer, a batch size of 128 and learning rate is set to 0.001.
In the event that the loss does not decrease for five epochs, the training epochs are stopped and the saved model is for prediction. The maximum number of training epochs is set to 50.

\subsection{Result and Analysis}\label{4C}
We compare our model with the baseline on the task of generating explanatory recommendations across three different datasets. We used two types of evaluation: explainability and textual quality. The overall results are presented in Tab.~\ref{tab3}, with the best performance value for each metric highlighted in bold. BLEU and ROUGE are presented as percentage values with the \% symbol omitted in the table, while the others are absolute values. 

Firstly, it is observed that our method consistently achieves the best performances in terms of explainability when compared to baseline methods across all three datasets. In particular, when looking over the USR and FCR metrics on TripAdvisor, our method outperforms others by more than double. For example, we identified that USR is boosted from 0.02 (Attn2Seq) to 0.57 (Ours), as well as FCR is surged from 0.02 (NRT) to 0.5 (Ours). We assume that the inadequate results obtained by Attn2Seq and NRT are attributed to their dependence on RNNs, which are susceptible to the well-known issue of long-term dependency problems. To avoid this issue, both PETER and PEPLER are embedded in the Transformer module where the feature within it can access all past features at any time step. Furthermore, the difference between PETER and PEPLER is whether the model has been pre-trained on the large language corpus, where latter conducted this process and achieved the better results than the former.

In the meantime, by simultaneously learning these two sub-tasks, the model can better comprehend user preferences and the explanations behind recommendation decisions.
This validates the effectiveness of our continuous prompt learning approach in using such knowledge to generate improved explanations.
In particular, when evaluating our model on the Amazon dataset, it consistently achieves the best performance across all metrics. This is because this dataset provides more ratings for the MF to reference, enabling better matching between users and items. 
This further supports the effectiveness of our model.

For the textual quality metric, we observed that the results of our approach has comparable performance with other SOTA models. Based on the Friedman statistic test, we identified that there is no significant difference between them across multiple datasets. This can be attributed to our utilization of a simple MF technique, where predictions are made by taking the dot product between user and item embeddings without involving additional parameters. 
However, there is a slight performance degradation when dealing with smallest dataset (TripAdvisor).
In general, our method ensures textual quality while significantly enhancing explainability.


\section{Conclusion}
In this paper, we propose a method that uses prompt and multi-task learning to generate a natural language sentence that explains why item $i$ is recommended to user $u$. 
Our approach involves inputting ID vectors directly as continuous prompts to GPT-2. 
We employ a joint training mechanism within a multi-task learning framework to optimize both the recommendation task and the explanation task. 
This enables us to better mine users' interests and needs, ultimately improving the recommendation effectiveness and user satisfaction. 
Additionally, we optimize the uncertain weights during training to determine the best weight values, improving the model's generalization ability and robustness.
Through experiments, we demonstrate that our solution is not only effective but also efficient. 
Our method achieves superior performance in terms of explainability and ensures stable text readability across three public datasets.

In future work, we would like to investigate a generic RS that incorporates our multi-task learning approach to address the challenge of dataset that has limited training samples.


{
\bibliographystyle{IEEEtran}
\bibliography{reference}
}

\end{document}